\documentclass[twocolumn,amssymb, showpacs, amsmath,superscriptaddress,prb]{revtex4}

\usepackage{graphicx}% Include figure files
\usepackage{dcolumn}% Align table columns on decimal point
\usepackage{bm}% bold math

\begin{document}

\title{Experimental evidence for a two-band superconducting state of NbSe$_2$ single
crystals}

\author {M. Zehetmayer}
\affiliation {Vienna University of Technology, Atominstitut, 1020
Vienna, Austria}
 \email{zehetm@ati.ac.at}
\author {H. W. Weber}
\affiliation {Vienna University of Technology, Atominstitut, 1020
Vienna, Austria}

\begin{abstract}
We report on measurements and a detailed analysis of the reversible magnetization of superconducting
NbSe$_2$ single crystals. By comparing the experimental data with Ginzburg Landau theory we show
that superconductivity in NbSe$_2$ cannot be explained by an anisotropic single-band, but by a
multi-band scenario. Applying a simple two-band model reveals the basic mixed-state parameters,
which are quite different in the two bands. We identify a strongly anisotropic band that determines
the properties at high magnetic fields, and a second almost isotropic band that dominates at low
fields. Our method is well suited for distinguishing anisotropic single-band from multi-band
superconductivity in various materials.
\end{abstract}

\pacs{74.25.Ha,74.25.Op,74.70.Ad}

\maketitle

\section{Introduction}

NbSe$_2$ is certainly one of the most intensively studied superconducting materials for several
reasons. First, its transition temperature ($T_{\text{c}}$) is about 7\,K and its upper critical
field not much larger than 4\,T perpendicular ($B^{\text{c}}_{\text{c2}}$, c-direction) and 12\,T
parallel ($B^{\text{ab}}_{\text{c2}}$, ab-direction) to the Nb planes.\cite{San95a} Accordingly,
most part of the superconducting phase diagram is accessible to experiment, in contrast to many
other materials. Furthermore, large high quality single crystals with almost negligible vortex
pinning effects can be grown. Introducing a small amount of disorder (e.g. by particle irradiation)
may lead to the emergence of the well known fishtail effect,\cite{Bha93a} which is still discussed
a lot by the superconductivity community. Moreover, NbSe$_2$ was the first material, in which
scanning tunneling microscopy was successfully employed for observing vortex cores or distributions
\cite{Hes89a,Hes90a} and it is still widely used for such investigations. Finally, the charge
density wave state, formed below about 33\,K,\cite{Mon75a} allows studying the effect of competing
order parameters, which is an important issue for high temperature superconductors.

Recently, NbSe$_2$ was suggested to be a two- or multi-band superconductor, which was confirmed by
several experiments (e.g. Refs.~\onlinecite{Yok01a,Boa03a,Rod04a,Fle07a,Hua07a,Hos09a}), but usually
an alternative interpretation of the results in terms of an anisotropic s-wave single-band scenario
could not be excluded. In this paper we provide further evidence for the two-band scenario. The
field dependence of the reversible magnetization - $M(B)$ - is compared with Ginzburg Landau theory,
showing that $M(B)$ cannot be reliably described by the anisotropic single-band scenario, but by
multi-band superconductivity. Evaluating the curves reveals the anisotropy and other superconducting
parameters which are not only temperature but also significantly field dependent.

More generally, following the discovery of superconductivity in MgB$_2$, multi-band scenarios were
announced for a lot of materials including Fe-pnictides, borocarbides, heavy fermions, and even
cuprates,\cite{Hun08a,Maz08a,Shu98a,Sey05a,Kha07a} etc. However, the experimental evidence was often
based on poor arguments (like, e.g., the observation of a temperature dependent anisotropy) and the
anisotropic single-band scenario could rarely be eliminated. Therefore, a method for distinguishing
between the two scenarios is clearly desirable, which is provided by our approach. At the same time,
all basic mixed state parameters can be assessed in this way.

\section{Experiment and evaluation}

Two NbSe$_2$ single crystals grown by a standard chemical vapor transport technique with sizes $a
\times b \times c \simeq 2.25 \times 2 \times 0.1$ mm$^3$ and $1 \times 1 \times 0.1$ mm$^3$ were
investigated. Both have a $T_{\text{c}}$ of 7.15\,K and a small transition width of about 0.1\,K.
The smaller crystal was mainly used for confirming the results of the larger one.

Measurements of the magnetic moment parallel to the applied field ($m$) as a function of the applied
field ($\mu_0 H_{\text{a}} \leq 7$\,T, $\mu_0 = 4\pi\times 10^{-7}$\,T m A$^{-1}$) were carried out
at temperatures from 2\,K to about 10\,K in our SQUIDs. The curve above $T_{\text{c}}$ (at 10\,K)
was subtracted from the superconducting signal to get rid of the temperature independent background
signal (which was non-significant in most cases). In all measurements, most parts of the
magnetization curves were reversible. Therefore, the reversible magnetization could be obtained
directly from $M = m/V$ ($V$ is the sample volume). In the case of irreversibility, the reversible
signal had to be calculated from the irreversible parts at increasing ($m_+$) and decreasing ($m_-$)
field via $M = (m_+ + m_-)/2 V$. Minor corrections to the field coming from the critical current
were considered numerically as described in Ref.~\onlinecite{Zeh09a}. Finally, the magnetic
induction $B$ was evaluated via $B = \mu_0 (H_{\text{a}} - DM + M)$, where $D$ is the
demagnetization factor of the sample, and $M(B)$ obtained. Fitting $M(B)$ from single-band Ginzburg
Landau (GL) theory to the experiment results in two parameters, the upper critical field
($B_\text{c2}$), which is the field where $M(B>0)$ vanishes for the first time, and $\kappa$, the GL
parameter. For simplicity, the GL results were approximated by simpler (but quite accurate)
equations given in Ref.~\onlinecite{Bra03a} (we used Eq. 23 of that reference leading to deviations
from the exact behavior by less than 3.5\%). We point out that other approximations or even
numerical solutions of the reversible magnetization $M(B)$ may be applied, which are obtained from
GL theory or other models.

\section{Results and discussion}

\subsection{Fitting the reversible magnetization by single- and two-band models}

\begin{figure}
    \centering
    \includegraphics[clip, width = 8.5cm]{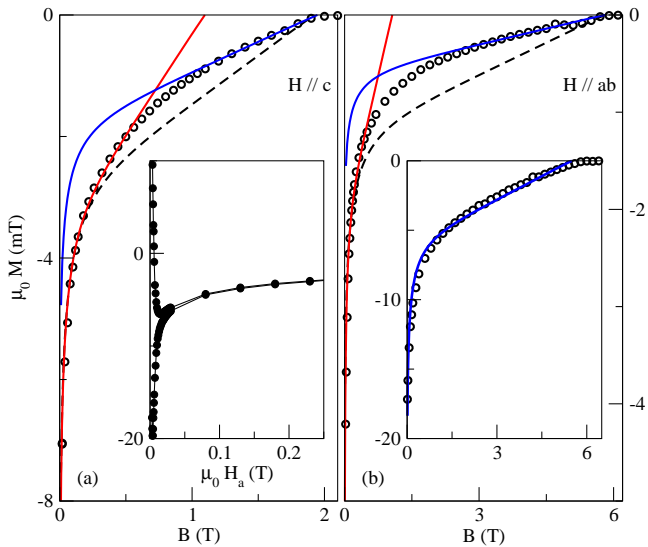}
    \caption{\label{fig1}(Color online) The reversible magnetization from experiment and GL theory.
The open circles illustrate the experimental data of NbSe$_2$ for $H_{\text{a}} \parallel c$ (panel
a) and $H_{\text{a}} \parallel ab$ (panel b) at 4.2\,K. The dashed lines present the best
single-band GL fits to the whole field interval, the solid lines to the low- or high-field region
only. The inset of panel (a) shows a measured hysteresis loop from which the reversible data has
been evaluated. The inset of panel (b) presents the reversible magnetization of a V$_3$Si single
crystal at 13.5\,K (open symbols) and the best GL fit (solid line).}
\end{figure}

Figure~\ref{fig1} presents results of the experiments and the fitting procedure. The full circles in
the inset of panel (a) show the low-field part of the measured magnetization loop $m/V$ for
$H_{\text{a}} \parallel c$ at 4.2\,K. A significant hysteresis occurs only at very low fields (below
about 20\,mT, which is $\sim 0.01 B_{\text{c2}}$) and this region was excluded from the reversible
curves presented by the open circles in the figure. The dashed lines show the best single-band GL
fit to the whole superconducting part (from about $0.01B_{c2}$ to $B_{c2}$) and therefore represent
the results we expect from a single-band superconductor. Here, $\kappa$ was the only fit parameter
while $B_{\text{c2}}$ was taken from the point, where the experimental curves reach zero. Enormous
differences between experiment and the best GL fits are evident (e.g. Fig.~\ref{fig1}) at both field
directions and all temperatures (2 - 6.6\,K). These deviations become even more apparent when
fitting the GL theory only to the high- or low-field region (in the latter case, also
$B_{\text{c2}}$ is fitted) as illustrated by the solid lines in Fig.~\ref{fig1}. We conclude that
NbSe$_2$ obviously does not behave like an anisotropic single-band superconductor.

The GL theory holds strictly only sufficiently close to the phase transition and $T_{\text{c}}$.
At low temperatures and far from $B_{c2}$ some deviations are expected, but usually GL theory is
quite successful  over the whole field and temperature range. Nevertheless, to clarify this point,
we fitted the single-band GL theory to magnetization curves of Nb and V$_3$Si, which are single-band
superconductors. In all cases the GL fit showed reliable agreement with experiment over
the whole field range, as illustrated, e.g., in the inset of Fig.~\ref{fig1}(b) for V$_3$Si. Fair
agreement between GL theory and $M(B)$ data was also found in the more anisotropic superconductors
\cite{Pog01a,Ach05a,Kar06a} YBa$_2$Cu$_4$O$_8$, Nd$_{1.85}$Ce$_{0.15}$CuO$_{4-\delta}$, and
YBa$_2$Cu$_3$O$_{7-\delta}$.

In the next step we analyzed possible two-band effects. In the anisotropic single-band scenario,
properties like the energy gap or the coupling strength vary within one band, while in the
multi-band model\cite{Suh59a} superconducting charge carriers exist in two (or more) electron bands
at the Fermi surface characterized by two (or more) sets of parameters. Two-band behavior can be
observed if two of the parameter sets are significantly different, e.g. for the upper critical field
or the anisotropy, and if both bands contribute substantially to the overall properties.
Additionally, interband coupling (pairing of electrons from different bands) is expected, which
leads, e.g., to one single transition temperature of both bands. Even quite small
interband coupling strengths would strongly mask the superconducting transition of the band with
the lower values of $T_\text{c}$ and $B_{\text{c2}}$, which, however, does not mean that the
properties follow the behavior of a conventional single-band superconductor.\cite{Nic05a} While the
temperature dependence of most quantities is often quite similar for anisotropic single-band and
two-band materials, the field dependence is rather different. To analyze two-band effects in the
experiment, we separated $M(B)$ into a high- and a low-field region and fitted both regions
independently by single-band GL theory (see solid lines of Fig.~\ref{fig1}).

In the high-field region,  good agreement was achieved in the range from about 0.6
$B^{\text{c}}_{\text{c2}}$ to $B^{\text{c}}_{\text{c2}}$ for $H_{\text{a}} \parallel c$ and from 0.5
$B^{\text{ab}}_{\text{c2}}$ to $B^{\text{ab}}_{\text{c2}}$ for $H_{\text{a}} \parallel ab$ at all
temperatures. This indicates that superconductivity of the second band is suppressed at those
high-fields and only the single-band properties of the band with the larger upper critical field
(indicated by the superscript '$\alpha$') survive. Some effects of interband coupling cannot be
excluded, but are expected to be small. Accordingly, the high-field GL fit is assumed to roughly
provide the single-band parameters of the $\alpha$-band (see Fig.~\ref{fig2} and Tab.~\ref{tab1}).

Turning to lower fields in Fig.~\ref{fig1}, the discrepancy between the experimental data and the
high-field single-band GL fit increases, which we ascribe to the influence of a second band with a
smaller $B_{\text{c2}}$ (superscript $\beta$). With decreasing field the $\beta$-band becomes more
dominant (particularly for $H_{\text{a}} \parallel ab$), but the contribution of $M^\alpha (B)$ does
not become negligible. Therefore, single-band behavior is not expected and the low-field GL fit only
provides effective parameters (index 'lf') of the overall superconducting behavior, although more
indicative of the $\beta$-band. The fit interval ranged from $0.05 B^{\text{c}}_{\text{c2}}$ to $0.3
B^{\text{c}}_{\text{c2}}$ for $H_{\text{a}} \parallel c$, which was the largest interval that
enabled good matching at all temperatures. The fit parameters are more affected by reducing the
interval than in the high-field regime, but the major findings are not concerned. To be consistent,
we  applied the same field interval in absolute values for $H_{\text{a}} \parallel ab$, which
roughly corresponds to $0.016 B^{\text{ab}}_{\text{c2}}$ to $0.1 B^{\text{ab}}_{\text{c2}}$.

To come closer to the single-band properties of the $\beta$-band, we subtracted the $\alpha$-band GL
fit from the experimental data. However, pure single-band behavior of the remaining $M(B)$
($M^\beta(B)$) is still not expected and indeed not observed due to interband coupling (which might
dominate, where $M^\beta(B)$ approaches zero, i.e. near $B_{\text{c2}}$ of the $\beta$-band) and
possible additional superconducting bands. Accordingly, the field interval, where single-band GL
theory matched well was similar as for the effective 'lf' regime before subtracting the
$\alpha$-band. In contrast to the 'lf' regime, however, the thermodynamic critical field
($B_\text{c}$, see Tab.~\ref{tab1}) of the $\beta$-band is almost independent of the field
orientation as expected for a single band (small deviations are ascribed to the interband coupling
and experimental uncertainties).

\subsection{Mixed-state properties and anisotropy}

\begin{figure}
    \centering
    \includegraphics[clip, width = 8.5cm]{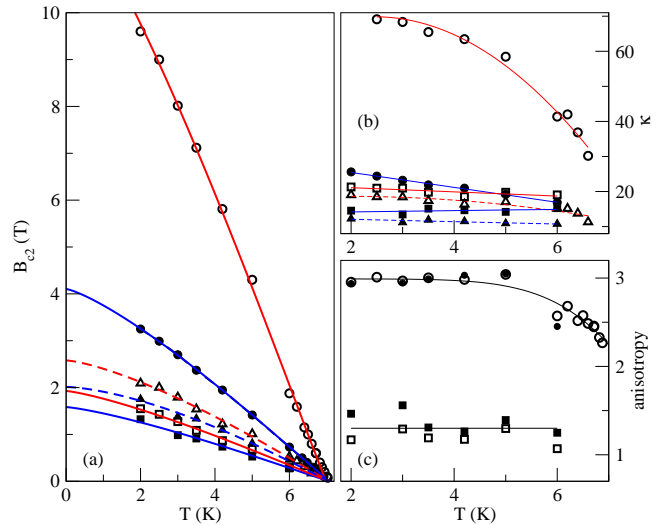}
    \caption{\label{fig2}(Color online) The superconducting parameters of NbSe$_2$ obtained from
evaluating the reversible magnetization curves. Panel (a) displays the upper critical fields
($B_{\text{c2}}$), panel (b) the Ginzburg Landau parameters ($\kappa$), and panel (c) the
anisotropies. The circles (with solid lines) present $\alpha$-band, the squares (with solid lines)
$\beta$-band, and the triangles (with dashed lines) effective low-field regime results. In panel (a)
and (b) the open symbols refer to $H_{\text{a}} \parallel ab$ and the full symbols to $H_{\text{a}}
\parallel c$, in panel (c) to the $B_{\text{c2}}$ (open) and the $\kappa$ (full) anisotropies. The
lines are guides for the eyes.}
\end{figure}

The results of the fitting procedure are illustrated in Fig.~\ref{fig2} and Tab.~\ref{tab1}.
$B_{\text{c2}}^{\text{c},\alpha}$ ($\alpha$-band) follows well the behavior of simple conventional
superconductors (full circles in Fig.~\ref{fig2}(a)). For extrapolating the data to 0\,K the model
$B_{\text{c2}}(T) = B_{\text{c2}}(0) [1 - (T / T_{\text{c}})^a]^b$ (lines in Fig.~\ref{fig2}(a)) was
applied which matches best for $a \simeq 1.3$, $b \simeq 1$, and
$B_{\text{c2}}^{\text{c},\alpha}(0\,\text{K}) \simeq 4.1$\,T. The same model leads to
$B_{\text{c2}}^{\text{ab},\alpha}(0\,\text{K}) \simeq 12.3$\,T ($H_{\text{a}} \parallel ab$). In the
latter case, we observe a positive curvature of $B_{\text{c2}}$ near $T_{\text{c}}$, while
$B_{\text{c2}}^{\text{c},\alpha}$ is linear in this regime. The corresponding GL parameters
($\kappa$, Fig.~\ref{fig2}(b)) decrease strongly with increasing temperature. Extrapolation gives
roughly $\kappa^{\text{c}, \alpha}(0\,\text{K}) \simeq 30$ and
$\kappa^{\text{c},\alpha}(6\,\text{K}) \simeq 17$ as well as $\kappa^{\text{ab},\alpha}(0\,\text{K})
\simeq 86$ and $\kappa^{\text{ab},\alpha}(6.6\,\text{K}) \simeq 30$.

The $\beta$-band upper critical fields are also well fitted by the above model leading to much
smaller results at 0\,K, namely $B_{\text{c2}}^{\text{c},\beta}(0\,\text{K}) \simeq 1.6$\,T and
$B_{\text{c2}}^{\text{ab},\beta}(0\,\text{K}) \simeq 1.9$\,T. Also the GL parameters decrease and
show a less pronounced temperature dependence, i.e., $\kappa^{\text{c},\beta}(0\,\text{K}) \simeq
15$ and $\kappa^{\text{c},\beta}(6\,\text{K}) \simeq 15$, $\kappa^{\text{ab},\beta}(0\,\text{K})
\simeq 21$, $\kappa^{\text{ab},\beta}(6\,\text{K}) \simeq 19$. The effective parameters (which
address the sum of the $\alpha$- and $\beta$-band) of the low-field regime deviate only slightly
from the $\beta$-band data (see  Fig.~\ref{fig2} and Tab.~\ref{tab1}).

The above results lead to the anisotropies $\gamma_{\text{Bc2}} = B^{\text{ab}}_{\text{c2}} /
B^{\text{c}}_{\text{c2}}$ and $\gamma_{\kappa} = \kappa^{\text{ab}} / \kappa^{\text{c}}$ illustrated
in panel (c) of Fig.~\ref{fig2}. At high fields ($\alpha$-band) and temperatures below $\sim 5$\,K,
$\gamma_{\text{Bc2}} \simeq 3$ and almost constant (open circles), i.e. close to the value usually
reported for NbSe$_2$,\cite{San95a} but drops to 2.3 at 7\,K. Such a temperature dependence
corresponds to the positive curvature of $B^{\text{ab}}_{\text{c2}} (T)$ near $T_{\text{c}}$ and is,
in general, a characteristic of both, two-band\cite{Shu98a} and anisotropic single-band\cite{Pit93a}
superconductivity. The anisotropy of $\kappa$ (full circles) is almost identical to
$\gamma_{\text{Bc2}}$ (open circles) in the high-field regime.
At lower fields the anisotropy of the GL properties is considerably reduced. We obtain $\gamma^\beta
\simeq 1.3$ ($\gamma_{\text{Bc2}} \simeq 1.2$, $\gamma_{\kappa} \simeq 1.4$) and $\gamma^\text{lf}
\simeq 1.35$ ($\gamma_{\text{Bc2}} \simeq 1.2$, $\gamma_{\kappa} \simeq 1.5$), thus almost isotropic
behavior. The small differences in $\gamma_{\text{Bc2}}$ and $\gamma_{\kappa}$ could result from the
larger experimental uncertainties in this regime.

In summary, the anisotropy of NbSe$_2$ changes strongly with the applied field from a large value
($\sim 3$) at high fields ($\gtrsim 0.5 B_{\text{c2}}$) to almost isotropic behavior at low fields
($\lesssim 0.3 B^{\text{c}}_{\text{c2}}$, $\lesssim 0.1 B^{\text{ab}}_{\text{c2}}$). It seems
impossible to explain this behavior by a single-band model, i.e. a two- or multi-band model is
needed. Our results suggest the existence of a strongly anisotropic ($\alpha$) band with a large
$B_{\text{c2}}$, and a second almost isotropic ($\beta$) band with a lower $B_{\text{c2}}$
(or, to be consistent with two-band theory: both bands may have the same $B_{\text{c2}}$ due to
interband coupling, but the properties of the $\beta$-band are strongly suppressed at high fields).
We further point out, that the anisotropies of $B_{\text{c2}}$ and $\kappa$ are equal or quite
similar, when acquired at the same field and temperature. Accordingly, this holds also for the
anisotropy of all further parameters, such as the magnetic penetration depth ($\lambda$) or the
coherence lengths ($\xi$). These anisotropies were often claimed to be quite different in the
two-band superconductor MgB$_2$. But detailed evaluations showed that the situation is actually the
same as in NbSe$_2$, i.e. all parameters have the same anisotropy.\cite{Zeh04a} The
misinterpretation in MgB$_2$ usually originated from measuring the anisotropies of different
properties at different applied fields, e.g. that of $\xi$ at high fields (from the $B_{\text{c2}}$
anisotropy) and that of $\lambda$ at low fields.

\begin{table}
\caption{\label{tab1} Summary of mixed-state parameters of single crystalline NbSe$_2$ at 0\,K in
the $\alpha$- and $\beta$-band and the effective low-field ('lf') regime. Minor inconsistencies with
the GL relations occur from extrapolating the experimental data to 0\,K.}
\begin{ruledtabular}
\begin{tabular}{lccclccc}
 &$\alpha$ & $\beta$ & lf &&$\alpha$ & $\beta$ & lf\\
\hline
$B^{\text{c}}_{\text{c2}}$ (T): & 4.1 & 1.6  & 2.0 &
$B^{\text{ab}}_{\text{c2}}$ (T): & 12.3 & 1.9  & 2.6 \\
$B_{\text{c1}}^{\text{c}}$ (mT): & 9.0 & 11.4 & 19.1 &
$B_{\text{c1}}^{\text{ab}}$ (mT): & 3.9 & 7.7 & 10.3 \\
$B_{\text{c}}^{\text{c}}$ (mT): & 98 & 67 & 112 &
$B_{\text{c}}^{\text{ab}}$ (mT): & 98 & 64 & 86 \\
$\lambda_{ab}$ (nm): & 265 & 215 & 165 &
$\lambda_c$ (nm): & 795 & 355 & 350 \\
$\xi_{ab}$ (nm): & 9 & 14 & 13 &
$\xi_c$ (nm): & 3 & 12  & 10 \\
$\kappa^c$: & 30 & 15 & 13 &
$\kappa^{ab}$: & 86 & 21  & 21\\
$\gamma$: & 3.0 & 1.3  & 1.4 &&&\\
\end{tabular}
\end{ruledtabular}
\end{table}

Further parameters were calculated by applying the well known GL relations (e.g.
Ref.~\onlinecite{Tin75a}), namely $\xi$, $\lambda$, $B_{\text{c}}$, and $B_{\text{c1}}$ (the lower
critical fields). The temperature dependence of these quantities was fitted by the same (general)
model as $B_{\text{c2}}$ and the results at 0\,K are listed in Tab.~\ref{tab1}. Additionally,
$B_{\text{c}}$ of the whole sample was obtained from integrating the reversible
magnetization curve,\cite{Tin75a} i.e. $-\mu_0 \int_0^{Hc2} M dH =
B_{\text{c}}^2 / (2 \mu_0)$, resulting in about 120 mT at 0\,K.

\subsection{Discussion}

Our results are consistent with most findings from literature. For instance, the Fermi level of
NbSe$_2$ is known to be crossed by three electron bands, one is related to Se-4p orbitals and shows
a rather three dimensional pancake like sheet, and the other two to Nb-4d orbitals each leading to
two rather two-dimensional (cylindrical) Fermi sheets.\cite{Joh06a} There is  general consensus
that a superconducting gap opens in both Nb bands. Investigations by ARPES\cite{Yok01a} at 5.3\,K
resulted in gap values of 0.9 - 1.0\,meV, but no gap was detected on the Se bands. On the other
hand, scanning tunneling spectroscopy\cite{Rod04a,Gui08a} revealed a broader range of gap values
from about 1.4 to 0.7 (or even 0.4) meV close to 0\,K. The band, on which the smaller gaps exist,
was not clearly identified, but it was argued, that the smaller gaps may open on the Se bands, which
would not necessarily contradict the ARPES measurements, since the smaller gap could be suppressed
near $T_{\text{c}}$ in a two-band model.\cite{Rod04a} It is plausible to associate the
$\alpha$-band, which has the higher $B_{\text{c2}}$, with the larger gap ($\Delta_\alpha$) and thus
the $\beta$-band with the smaller gap ($\Delta_\beta$). The fact that the superconducting properties
of the $\beta$-band are nearly isotropic may favor assigning $\Delta_\beta$ to the more three
dimensional Se-band. On the other hand, ab-initio calculations revealed that the electron density of
states of the Se band at the Fermi level - $z(0)$ - is only about 5\,\% of the overall
value,\cite{Joh06a} which makes a substantial contribution of the Se bands to the superconducting
behavior questionable. Moreover, applying the BCS relations for the condensation energy density
$E_\text{c} = -z(0) \Delta^2 / 2 = -B_\text{c}^2 / (2 \mu_0)$ with $B_\text{c}$ from
Tab.~\ref{tab1}, results in similar $z(0)$ values for both bands, as was actually found for the two
Nb bands in Ref.~\onlinecite{Joh06a} and thus favors a scenario, in which the two Nb bands are
responsible for the two-band effects. It may be concluded that some microscopic aspects remain
unclear and need further investigations. We note further that our results are in fair agreement with
those from a specific-heat study.\cite{Hua07a} Fitting the field dependence of the specific heat
data by a simple two-band model led to similar upper critical fields and anisotropies of the two
bands as in our study.

Finally, we wish to mention the remarkable qualitative resemblance of the
two-band properties in NbSe$_2$ and MgB$_2$, for which a similar analysis\cite{Zeh04a} also
demonstrated a highly anisotropic band ($\sigma$-band) being dominant at high magnetic fields and a
quite isotropic low-field band ($\pi$-band), with similar values for the upper critical fields and 
the anisotropies, but a larger transition temperature (of almost 40\,K). We point out that 
the mixed-state parameters obtained by our approach in MgB$_2$ are in good agreement with results
derived from other methods, even for the band with the lower $B_\text{c2}$.

There is also cumulative evidence for multi-band superconductivity in the family of the Fe-pnictide
superconductors (e.g. Refs. \onlinecite{Hun08a,Maz08a}). Contrary to NbSe$_2$ and MgB$_2$, however,
the anisotropy of the upper critical field (i.e., the high-field value of $\gamma$) seems to be
smaller than that at low fields \cite{Kha09a,Wey09a}(i.e. the anisotropy of the penetration depth or
of the lower critical field). This would indicate that the band with the larger upper critical
field is less anisotropic. More recent results suggest, however, that a proper description of
experimental data is only possible by taking three or more bands into account.\cite{Umm09a,Kim10a}
Further insight might be gained when reversible magnetization curves over the whole field range
become available for these materials.

\section{Summary}

In summary, we have shown that the reversible magnetization of NbSe$_2$ cannot be reliably explained
by a standard anisotropic single-band scenario. Applying a simple two-band model reveals the
anisotropy and other mixed-state parameters as a function of magnetic field and temperature. We
found a strongly anisotropic band ($\gamma \simeq 3$) that dominates at high fields and a second
almost isotropic band relevant only at low fields. The method can be applied to other materials to
distinguish between single-band and multi-band superconductivity and to acquire at the same time all
basic mixed-state parameters.

\begin{acknowledgments}
The authors thank P. H. Kes for providing samples and A. Kremsner and C. Trauner for carrying out
some of the measurements. This work was supported by the Austrian Science Fund under contract 21194.
\end{acknowledgments}

\end{document}